\documentclass[reprint,
superscriptaddress,
prl,
]{revtex4-1}

\usepackage{xcolor}
\usepackage{graphicx}
\usepackage{dcolumn}
\usepackage{bm}
\usepackage{makecell}

\usepackage{textcomp} 

\usepackage{gensymb}

\begin{document}
\title{Terahertz Nonlinear Optical Response of Water Vapor}

\author{Payman Rasekh}
\email[]{p.rasekh@gmail.com}
\affiliation{School of Electrical Engineering and Computer Science, University of Ottawa, Ottawa, ON, Canada.}

\author{Akbar Safari}
\email[]{akbar.safari@gmail.com}
\affiliation{Department of Physics, University of Ottawa, Ottawa, ON, K1N 6N5, Canada.}

\author{Murat Yildirim}
\affiliation{Department of Physics, University of Ottawa, Ottawa, ON, K1N 6N5, Canada.}

\author{Ravi Bhardwaj}
\affiliation{Department of Physics, University of Ottawa, Ottawa, ON, K1N 6N5, Canada.}

\author{Jean-Michel M\'enard}
\affiliation{Department of Physics, University of Ottawa, Ottawa, ON, K1N 6N5, Canada.}

\author{Ksenia Dolgaleva}
\affiliation{School of Electrical Engineering and Computer Science, University of Ottawa, Ottawa, ON, Canada.}
\affiliation{Department of Physics, University of Ottawa, Ottawa, ON, K1N 6N5, Canada.}

\author{Robert W. Boyd}
\affiliation{School of Electrical Engineering and Computer Science, University of Ottawa, Ottawa, ON, Canada.}
\affiliation{Department of Physics, University of Ottawa, Ottawa, ON, K1N 6N5, Canada.}
\affiliation{Institute of Optics, University of Rochester, Rochester, New York, 14627, USA.}

\vspace{3mm}

\date{\today}

\begin{abstract}
We report on the nonlinear spectroscopy of water vapor at THz frequencies. Atmospheric water vapor has a rich spectrum with several strong resonances at frequencies below 3~THz, falling within the range of operation of most existing THz sources. We observe an extremely large nonlinear response to THz radiation at the positions of these resonances. Using the optical Kerr model for the nonlinear response, we estimate a minimum nonlinear refractive index of the order of $10^2$~$\text{m}^2/\text{W}$.
Our results provide insight into the energy levels of the water molecule and give a more accurate picture of its response to electromagnetic radiation, paving the way to more accurate THz spectroscopy, imaging and sensing systems, and thereby facilitating future emerging THz technologies.
\end{abstract}

\maketitle
Water vapor pervades the atmosphere of the Earth and exhibits absorption in a broad frequency range of the electromagnetic spectrum, associated with rotational, as well as inter- and intra-molecular vibrational transitions~\cite{Deepak1980,SLOCUM201349}. The absorption spectrum of water vapor reflects its molecular structure and, therefore, can be used as a spectroscopic fingerprint for its detection and identification. This spectrum is, however, very complex and highly rich in resonances, which have been the center of intense spectroscopic investigations for many decades~\cite{Horvath1993,Mecke1933,Elsasser1938,Savolainen20402}. Moreover, atmospheric water vapor is responsible for almost all the spectral modifications of electromagnetic radiation from 0.5 to 20 THz~\cite{Afsar1978}. It is, therefore, crucial to gain a better understanding of these resonances, especially in the context of high field excitation, due to an increasing demand for high power sources to enable numerous emerging THz applications. 

Indeed, THz-based systems are finding a niche in fields such as information and communication technology~\cite{Nagatsuma2011}, spectroscopy and imaging~\cite{Dong2017,Fischer2016,Zimdars2005}, ultrafast control~\cite{Bowlan2017}, as well as biomedicine~\cite{Brun2010,Tuchin2018}. In addition, long-path THz time-domain spectroscopy in the open air is realized to demonstrate the potential of line-of-sight THz communications~\cite{Grischkowsky2015,Grischkowsky2017}, THz sensing~\cite{Kurz2000}, monitoring pollutants and dangerous gases~\cite{Yasui2016}, non-destructive evaluation~\cite{Blackshire2008}, global environmental monitoring~\cite{Kaushik2012}, and THz imaging through the fog and smoke~\cite{Grishkowsky2015Fog}, which often require propagation of intense THz pulses through the atmosphere. Therefore, studying the nonlinear response of water vapor in the THz domain not only provides insight into the inter-molecular structure of water vapor but also represents great importance for practical applications.

With the coherent generation and detection of THz radiation, it is possible to measure not only absorption spectra with high spectral resolution, but also the dispersive response of water vapor simultaneously~\cite{Grischkowsky2012}. With recent developments in coherent THz radiation sources, phase-locked THz pulses with higher intensities are becoming routinely accessible~\cite{HeblingPhonon2008}. As a consequence, THz science has been engaged in studying the nonlinear response of different materials in the THz region of spectrum. For instance, the nonlinear response of free electrons to an intense single-cycle THz pulse was reported in~\cite{Turchinovich2012}. Moreover, THz-induced carrier multiplication via impact ionization was reported in~\cite{Tarekegne2015,Tani2012,Lange2014}, and THz saturable absorption and higher-harmonic generation by hot electrons was demonstrated in~\cite{Gaal_GaAs,Chai2018,Hwang2013,Razzari2009,Hafez2018,Schubert2014}.
In particular, it has been shown that a phonon-induced THz Kerr effect can result in larger nonlinear responses than that of the optical Kerr effect~\cite{HeblingPhonon2008,Ksenia2015,Payman2020}.
However, unlike solids and liquids, gaseous materials such as water vapor can have sharp resonances in the THz range, and the nonlinear THz response of these transitions has not been studied to date.

In this Letter, we report on the nature of nonlinear interactions of THz pulses with atmospheric water vapor at different THz pulse intensities. First, we measure the linear response of water vapor and explain our results by comparing them with theoretical predictions. Then, we perform nonlinear THz time-domain spectroscopy (THz-TDS) to show how increasing the electric field intensity of the THz radiation modifies the absorption and dispersion of the vapor. In contrast to two-level atomic systems, where the absorption decreases with increasing field intensity, we observe an increase in absorption for some of the transitions. The observed reverse-saturation of absorption is explained based on stepwise multi-photon transitions. Furthermore, we show that the THz field experiences an extremely large nonlinear refractive index at frequencies near to the resonances of water molecules.

\begin{figure*}[ht!]                    
\begin{center}
\includegraphics[width=1\textwidth]{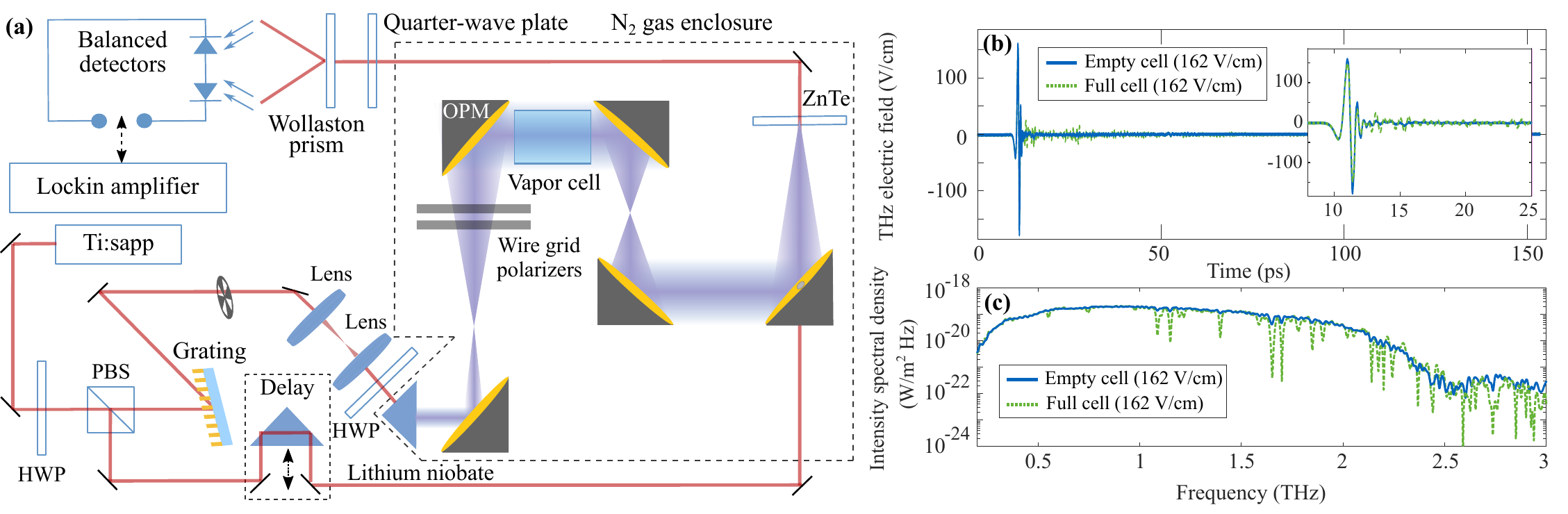} 
\end{center}
\caption{(a) Experimental setup diagram for THz generation and detection. The vapor cell is placed between two parabolic mirrors where the THz beam is collimated. A half-wave plate (HWP) and a polarizing beam splitter (PBS) are used to split the pump and probe beams. OPM is off-axis parabolic mirror. (b) THz field plotted as a function of time with and without water vapor; these plots show the transient response of the overall system. The dashed line shows the THz electric fields in the presence of the water vapor, while the solid line shows the reference signals collected in the absence of the vapor. The inset in (b) enlarges the main peaks and the trailing oscillations. (c) Intensity spectral density with and without water vapor obtained from Fourier transform of the time-domain signals of part (b). The THz spectrum extends to 2.5~THz, but we focus on the resonances between 1 and 1.5~THz where the signal-to-noise ratio of the source is maximum.}
\label{Setup} 
\end{figure*} 

The THz field is generated using optical rectification in a lithium niobate crystal with the tilted-pulse-front technique~\cite{HeblingGeneration2008}. In the generation arm [Fig.~\ref{Setup}~(a)], we use an 800-nm, 45-fs Ti:sapphire pulsed laser beam that diffracts from a grating that is used to introduce the required tilt to the intensity front of the pulse and to fulfill the phase matching condition of the nonlinear-induced polarization in the lithium niobate. Off-axis gold parabolic mirrors are used to collimate and expand the THz radiation. After the second cylindrical lens, we spatially filter the diffracted near-infrared laser beam to remove the unwanted diffraction orders and to generate a THz beam with a smooth near-Gaussian transverse profile. A pair of free-standing wire-grid polarizers is used after the first parabolic mirror to control the intensity of the THz beam without introducing dispersion to the transmitted field. 

In the detection arm, both the THz beam transmitted through the vapor cell and a near-infrared probe pulse overlap at the focal position into a $200$-$\mu\text{m}$ thick ZnTe crystal. In the absence of a THz electric field, a quarter-wave plate converts the linear polarization of the probe beam into circular polarization. When the THz beam co-propagates with the probe beam, the THz field modifies the optical refractive index seen by the probe pulse by means of the linear electro-optic (Pockels) effect, thereby inducing an ellipticity to the probe's polarization. We measure this ellipticity as a function of time delay with respect to the THz pulse in order to determine the time dependence of the THz field. A Wollaston prism along with a pair of balanced photodetectors and a lock-in amplifier allow for a linear detection of the induced birefringence at the chopping frequency of the probe. The peak of the electric field at the detection crystal is calculated based on the modulation of the balanced photodetectors using the procedure given in~\cite{Blanchard2007}, and is then re-scaled proportional to the beam diameters to obtain the peak electric field at the location of the vapor cell. We estimate the maximum peak value of the THz electric field to be $\approx$~2.7~kV/cm at the location of the vapor cell. Moreover, the THz part of the setup is enclosed and purged with dry nitrogen to prevent unwanted absorption from water vapor.

The water vapor cell is a cylinder with a length of 8~cm and a diameter of 5.5~cm. To avoid absorption and reflection of the THz field, ultra-thin cling wraps are used as the windows of the cell~\cite{Grishkowsky1989}. The vapor cell is placed in the collimated arm of the THz field and filled with dry nitrogen under atmospheric pressure. A few drops of distilled water are placed inside the cell to provide the required water vapor. A heating wire wrapped around the cell and a thermocouple are used to keep the temperature of the vapor cell at $T=309^\circ\,\text{K}$. To determine the effect of water vapor on the transmitted field, we collect the data with and without the water vapor. We repeat the measurements at different THz field amplitudes by rotating the first polarizer while keeping the second one fixed to maintain a constant polarization state.
 
Figure~\ref{Setup}~(b) shows the time-domain signals for the lowest THz amplitude corresponding to $162$ $\text{V}/\text{cm}$, where a linear response of water vapor is expected. The effect of the sharp resonances of the water vapor appears as the trailing oscillations due to the free induction decay of the resonances. A long scan time of 155~ps is used during the measurements in order to collect as much spectral information as possible. The spectral density of the THz field is shown in Fig.~\ref{Setup}~(c) where the fast Fourier transform (FFT) of the temporal shape is displayed. The zero-padding technique~\cite{LyonsBook} is applied on the temporal signal to achieve a smoother interpolation in the frequency domain.

For this study, we focused our attention on the spectral region between 1 and 1.5~THz where water vapor exhibits six strong resonances and where our THz system features optimal signal-to-noise ratio. The linear susceptibility of the vapor due to these six resonances is given by
\begin{eqnarray}                      
\chi = \chi_R + i \chi_I = \sum_{j=1}^{6} \frac{c^2}{2\pi^2\nu_j} \frac{\nu_j-\nu+i\gamma_j}{\gamma_j^2+(\nu_j-\nu)^2}\,n_w\,S_j,
\label{eq:Susc} 
\end{eqnarray}
where $c$ is the speed of light in cm/s, $n_w=1.389\times 10^{18}$~cm$^{-3}$ is the number density of the water molecules and $\nu_j$ is the resonant frequency of the $j$th transition. The linewidth of each transition depends on the temperature $T$ according to
\begin{eqnarray}                      
\gamma_j = c \left( \frac{296}{T} \right)^{a_j} \left( \gamma_s P_w + \gamma_f P_f\right),
\label{eq:gamma} 
\end{eqnarray}
where $a_j$ is the temperature exponent for the air broadening, and $P_w$ and $P_f$ are the partial pressures of the water vapor and the nitrogen gas in the cell. The self-broadening and foreign gas broadening coefficients of these six transitions are approximately the same and are found to be $\gamma_s=0.5$ and $\gamma_f=0.1$~cm$^{-1}\,$atm.$^{-1}$, respectively.  The required parameters are extracted from HITRAN database~\cite{hitran2017} and \cite{HOSHINA2008} and are listed in Table~\ref{table1}. 

\begin{table}[!tp]		    
\caption{Transition parameters of the six strongest resonances of water vapor molecules in the spectral range between 1 and 1.5~THz. The vapor cell is at $T=309^\circ\,\text{K}$ and contains one atmosphere of nitrogen gas. $\nu_j$ and $\gamma_j$ are the central frequency and the linewidth, $a_j$ is the air broadening exponent, $A_j$ is the Einstein $A$ coefficient, $g_j''$ is the degeneracy factor of the excited state, and $S_j$ is the spectral line intensity of the $j$th transition.}
\centering
\begin{tabular}{c|c|c|c|c|c|c}
\hline
\hline
$j$ & $\nu_j$ &  $a_j$  &  $\gamma_j$ &     $A_j$    &  $g_j''$   &  $S_j\times 10^{20}$  \\
  &  (THz)   &       & (GHz)     & (s$^{-1}$) &        & (cm/molecule) \\
\hline 
1 &  1.0974  &  0.78   &  3.91   &   0.0164  &  21    &  15.18    \\
2 &  1.1133  &  0.79   &  3.40   &   0.0184  &  3    &  4.521    \\
3 &  1.1629  &  0.78   &  3.70   &   0.0227  &  21    &  16.67   \\
4 &  1.2076  &  0.81   &  3.49   &   0.0283  &  9     &  5.308   \\
5 &  1.2288  &  0.76   &  3.73   &   0.0187  &  5     &  4.421   \\
6 &  1.4106  &  0.80   &  3.70   &   0.0426  &  33   &  13.87   \\
 \end{tabular}
\label{table1}
\end{table}

The spectral line intensity for each transition $S_j$ is a function of temperature. In thermodynamic equilibrium, the population distribution between the energy levels is governed by Boltzman statistics and changes with temperature.  Therefore, the spectral line intensity for each transition can be calculated from
\begin{eqnarray}                      
S_j = \frac{A_j c}{8\pi \nu_j^2} \frac{g_j'' e^{-h\nu_j/k_BT} \left( 1-e^{-h\nu_j/k_BT} \right)}{\sum_k g_k e^{-h\nu_j/k_BT}},
\label{eq:S} 
\end{eqnarray}
where $A_j$ is the Einstein $A$ coefficient and $g_j''$ is the degeneracy factor of the excited state (Table~\ref{table1}). $h$ and $k_B$ are Planck and Boltzman constants, respectively. The sum in the denominator of Eq.~(\ref{eq:S}) represents the total internal partition sum and can be found in HITRAN database for water molecules~\cite{hitran2017}. The last column of Table~\ref{table1} shows the spectral line intensities at $T=309 ^{\circ}\,\text{K}$ calculated from Eq.~(\ref{eq:S}).

\begin{figure}[!tp]                    
\begin{center}
\includegraphics[width=1\columnwidth]{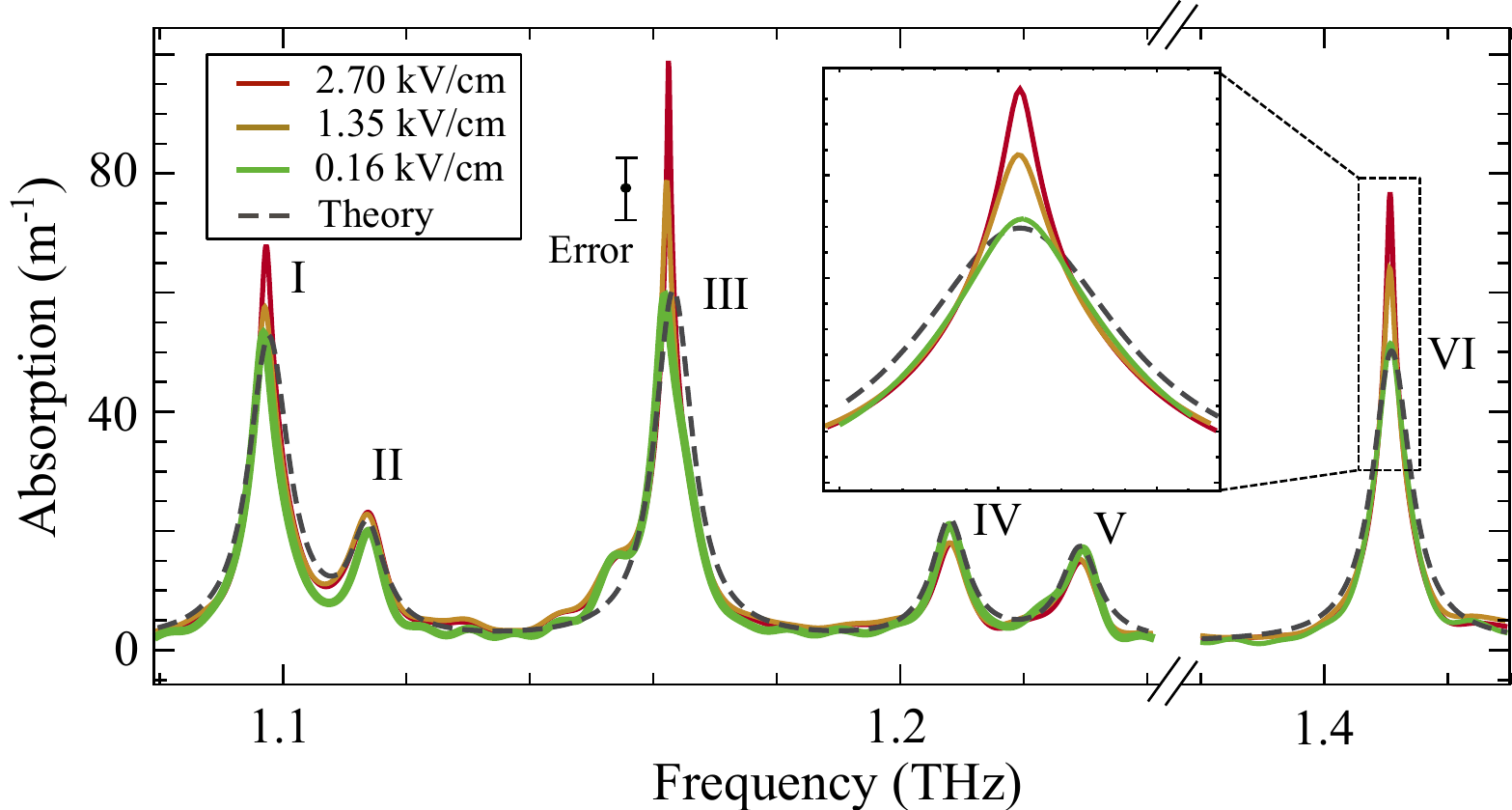} 
\end{center}
\caption{Absorption coefficient of water vapor at temperature $T=309^\circ\,\text{K}$. The absorption peaks are numbered from I to VI, and the corresponding labels are shown next to each peak. While the absorption of the lowest-intensity signals fits well with the linear model, the absorption at the resonances I, III and VI increases with the intensity.}
\label{fig:absorption} 
\end{figure}

In Fig.~\ref{fig:absorption}, we plot the calculated (linear response) and experimentally measured absorption coefficients. 
We notice that for the lowest intensity signal, the absorption coefficient fits well with the linear theoretical model. However, as the strength of the field increases, there is a consistent increase in the absorption at the location of the resonances I, III and VI. This change is an indication of the  nonlinear process associated with stepwise multi-photon absorption.

\begin{figure}[ht]                     
\centering 
\includegraphics[width=0.7\columnwidth]{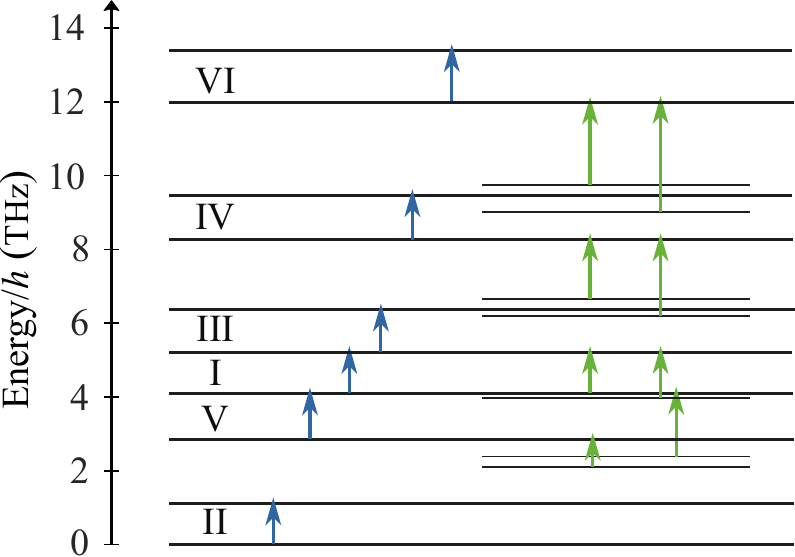} 
\caption{Energy level diagrams for the rotational transitions of water vapor where the lower and upper states of each transition under study is shown. The shorter lines on the right show the degenerate states at which an upper state of one transition coincides with the lower state of some other transition. Except for the second resonance, where the lower energy state coincides with the ground state, there is at least one degenerate transition for other resonances that pump the main transition.}
\label{fig:energy} 
\end{figure}

With a two-level-system approximation, one expects to see saturation of the transition, and thus, a decrease in absorption as the intensity increases. However, Fig.~\ref{fig:absorption} clearly shows that absorption of water molecules at some resonances increases with intensity, which indicates that two-level approximation is not valid here. In Fig.~\ref{fig:energy}, we plot the rotational energy levels of water molecules relevant to the six transitions in our study. 
The blue arrows show the transitions corresponding the the six resonances shown in Fig.~\ref{fig:absorption}. As can be seen, the transition V populates the ground state of transition I, which consecutively, populates the ground state of transition III. Therefore, a large increase in the absorption of the third transition is expected as its ground state population increases with THz field intensity. Moreover, the THz field has a broad spectrum covering the frequency window from $0$ to $3$~THz which excites many other transitions as can be seen in Fig~\ref{Setup}(c). The excited state of some of these transitions are indeed the ground state of one of the six transitions in our study. These transitions are shown by the green arrows in Fig.~\ref{fig:energy}. For example, the ground state of the sixth transition is pumped by two other transitions at $2.2$~THz and $2.97$~THz.
\begin{figure}[ht]                    
\begin{center}
\includegraphics[width=1\columnwidth]{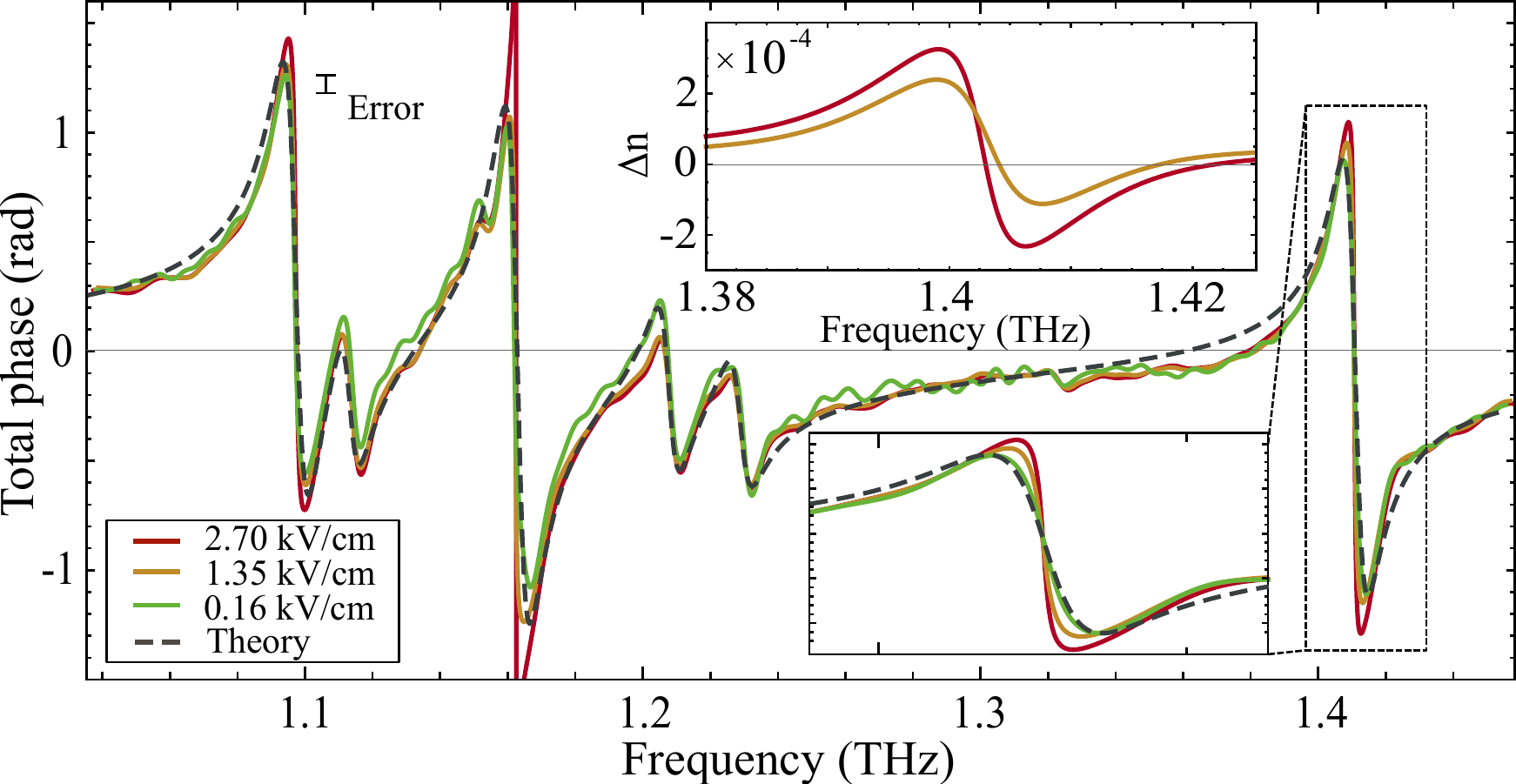} 
\end{center}
\caption{Overall phase of the transmitted THz field, including linear and nonlinear contributions, as a function of frequency. As the intensity of the field increases, the amplitude of the phase in the vicinity of the resonances gets larger. The top inset shows the amount of the refractive index change for the resonance at 1.4~THz.}
\label{fig:phase} 
\end{figure} 

Furthermore, since we use a coherent detection technique, we are able to retrieve the spectral phase information of the THz field from the Fourier analysis. Fig.~\ref{fig:phase} shows the phase of the THz field after the vapor cell as a function of frequency. It can be seen from the graph that the theoretical linear model fits well with the phase of the lowest-intensity signal, which is expected to be linear. Nonlinear response of the material can be found where the higher-intensity signal acquires additional phase shift,  specifically, at the location of the water vapor resonances. This nonlinear phase shift can be understood again from stepwise multi-photon transitions and pumping of the ground states of different transitions by the broadband THz pulse, as explained earlier.

By subtracting the phase of the lowest-intensity signal from those of the higher-intensity signals, we find the induced nonlinear phase shift $\Delta \phi$ of the THz field. Then, the nonlinear change in the refractive index can be calculated from~\cite{Boyd2008book}
\begin{equation}
\Delta n = \frac{\Delta \phi}{k\,L},
\label{eq:5}
\end{equation}
where $L$ is the length of the cell and $k$ is the wavenumber. This change in the refractive index is shown for the last resonance in the top inset of Fig. \ref{fig:phase}.

To the lowest order of approximation, the nonlinear change in the refractive index is often modeled by $\Delta n = n_2\,I$, where $I$ is the intensity of the field and $n_2$ is called the optical Kerr coefficient or nonlinear refractive index. However, in our work, $\Delta n$ arises from stepwise multi-photon transitions for which the spectral intensity of the THz field varies [see Fig. \ref{Setup}(c)]. Thus, one cannot assign a single intensity to each resonance that contributes to $\Delta n$ independently. Therefore, the nonlinear response of the vapor cannot be characterized by a simple Kerr coefficient, unambiguously. Nevertheless, if we consider the maximum spectral density for each resonance, we find a maximum intensity in a linewidth of the resonances to be around $10^{-6}$~$\text{W}/\text{m}^2$, and thus, we can estimate a minimum Kerr coefficient of the order of $10^2$~$\text{m}^2/$W. 

In conclusion, we observed an extremely large nonlinear response of water vapor in the THz regime. We attribute this large nonlinearity to stepwise multi-photon transitions in water molecules. Time-domain spectroscopy has been used widely to measure the absorption spectrum of water vapor in the THz regime. However, our results show that optical pumping of subsequent transitions can modify the observed spectrum even at low intensities. A primary obstacle to many THz applications is the attenuation of the field by water vapor in the atmosphere. A naïve approach to overcome this problem is to increase the THz intensity to saturate the transition and reduce the absorption. However, our study shows that, as the intensity increases, the absorption of many resonant lines increases as well. Moreover, nonlinear spectroscopy provides valuable information on the transitions and the energy levels of the molecules. Thus, our work paves the way for a better understanding of water vapor molecular structure, to validate the rotational energy levels of water molecules and to study the role of water clusters in the absorption spectrum~\cite{IUPAC2013,Carlon1979,Carlon1981,JOHNSON20086037,DAI2019277}.

\subsection{Acknowledgement}
This work was supported by the Canada Excellence Research Chairs program, Canada Research Chairs program and the National Science and Engineering Research Council of Canada (NSERC) Discovery and Strategic programs. RWB also thanks the US Army Research Office and the Office of Naval Research for their support.

\bibliographystyle{aipauth4-1}

\end{document}